\documentclass[journal,9pt]{IEEEtran}
\usepackage{cite}
\usepackage{amsmath,amssymb,amsfonts}
\usepackage{algorithmic}
\usepackage{graphicx}
\usepackage{textcomp}
\usepackage{multirow}
\usepackage{subfigure}
\usepackage{color}

\def\BibTeX{{\rm B\kern-.05em{\sc i\kern-.025em b}\kern-.08em
    T\kern-.1667em\lower.7ex\hbox{E}\kern-.125emX}}
\begin{document}
\title{{\fontsize{24}{26}\selectfont{{}}}\break\fontsize{16}{18}\selectfont
Solution of the Electric Field Integral Equation Using a Hybrid Quantum-Classical Scheme: Investigation of Accuracy and Efficiency}
\author{Rui Chen, Teng-Yang Ma, Meng-Han Dou, and Chao-Fu Wang
\thanks{This work received support from the National Natural Science Foundation of China (NSFC) under Grant 62201264 and Grant 62331016, from the Fundamental Research Funds for the Central Universities under Grant 30924010207, from the Fund Program for the Scientific Activities of Selected
Returned Overseas Professionals in Shanxi Province under Grant 20240063. \textit{(Corresponding author: Chao-Fu Wang.)} }
\thanks{Rui Chen and Chao-Fu Wang are with the School of
Microelectronics, Nanjing University of Science and Technology, Nanjing
210094, China (e-mail: rui.chen@njust.edu.cn; cfwang@ieee.org). }
\thanks{Teng-Yang Ma and Meng-Han Dou are with the Origin Quantum Computing Technology (Hefei) Co., Ltd., Hefei 230088, China (e-mail: mtc@originqc.com; dmh@originqc.com).}}

\maketitle

\begin{abstract}
 Conventional classical solvers are commonly used for solving matrix equation systems  resulting from the discretization of surface integral equations (SIEs) in computational electromagnetics (CEM). However,  the memory requirement would become a bottleneck for classical computing as the electromagentic problems become much larger. As an alternative, quantum computing has a natural “parallelization” advantage with much lower storage complexity due to the superposition and entanglement in quantum mechanics. Even though several quantum algorithms have been applied for the SIEs-based methods in the literature, the size of the matrix equation systems solvable using them is still limited. In this work, we use a hybrid quantum-classical scheme to solve the electric field integral equation (EFIE) for analyzing electromagentic scattering from three-dimensional (3D) perfect electrically conducting objects with arbitrary shapes in CEM for the first time. Instead of directly solving the original EFIE matrix equation system using the quantum algorithms, the hybrid scheme first designs the preconditioned linear system and then uses a  double-layer iterative strategy for its solution, where the external iteration layer builds subspace matrix equation systems with smaller dimension and the internal iteration layer solves the smaller systems using the quantum algorithms. Two representative quantum algorithms, Harrow-Hassidim-Lloyd (HHL) and variational quantum linear solver (VQLS), are considered in this work, which are executed on the quantum simulator and quantum computer  platforms. We present the theoretical time complexity analysis of the hybrid quantum-classical  scheme and perform numerical experiments to investigate the accuracy and efficiency of the hybrid scheme. The results show that the computational complexity of the hybrid VQLS-classical scheme is lower than the conventional fast solvers in classical computing, which indicates the hybrid scheme is more promising for analyzing large-scale electromagnetic problems.

\end{abstract}


\section{Introduction}
\label{sec:introduction}
In the last several decades, numerical methods based on surface integral equations (SIEs) are widely used for analyzing time-harmonic electromagentic scattering from  perfect electrically conducting (PEC) objects in computational electromagnetics (CEM) \cite{Notaroš} . 

One necessary step of the SIEs-based methods is solving dense matrix equation systems resulting from the discretization of SIEs using classical direct solvers, iterative solvers, and their fast versions \cite{Chew, Bleszynski, Seo, Ostrowski}. Even though the time complexity of the solution of the systems using these solvers has been reduced from $\mathcal{O}(N^3)$ to $\mathcal{O} (N \mathrm{log} N)$, it still has the memory complexity of $\mathcal{O} (N \mathrm{log} N)$, where $N$ is the number of unknowns. Therefore, as the electromagentic scattering problems become larger (i.e., as $N$ keeps increasing), the memory requirement would become a bottleneck for classical computing (using ``bit" as the basic information element).  

Different from classical computing, quantum computing (using ``qubit" to process information) has a natural ``parallelization" advantage due to the superposition and entanglement in quantum mechanics \cite{Hidary}. It can simultaneously deal with $N$ quantum states with the qubit storage complexity of $\mathcal{O} (\mathrm{log} N)$ for the solution of the matrix equation systems. Obviously, quantum computing is a promising alternative for solving large-scale electromagnetic scattering problems. 

Recently, quantum computing has been applied for the SIEs-based methods in the literature.  In \cite{Cai}, a quantum algorithm implemented by the dense Hamiltonian simulation technique is proposed for the method of moments (MoM) in CEM. In \cite{Phillips2021}, the Harrow-Hassidim-Lloyd (HHL) quantum algorithm is used for the solution of  matrix equations for analyzing two-dimensional (2D) electromagnetic problems. In \cite{Song-B}, a  variational quantum algorithm is developed to enhance the subentire-domain basis functions method for analyzing electromagnetic properties of finite periodic planar structures. However, to the best of the authors' knowledge, the size of the matrix equation systems solvable using these quantum algorithms is still limited, which prevents the further application of them for solving large-scale electromagnetic problems for the SIEs-based methods. 

In \cite{Ye, Chen-Z}, a hybrid quantum-classical scheme, which does not suffer from this issue, has been proposed to solve differential equations in computational fluid dynamics (CFD). This scheme uses classical subspace methods \cite{Saad2003} to map the original linear systems into smaller subspace ones and then solves them using the quantum algorithms. It enables the simulation of large-scale CFD problems based on near-term quantum computers.

In this work, we apply the hybrid quantum-classical scheme to solve the electric field integral equation  (EFIE) for analyzing electromagentic scattering from three-dimensional (3D) PEC objects with arbitrary shapes in CEM for the first time. Instead of directly solving the original EFIE system using the quantum algorithms as reported in the literature, the hybrid scheme first designs a preconditioned linear system and then uses a double-layer iterative solution strategy, where the external iteration layer maps the preconditioned  system into subspace linear systems with the smaller dimension and the internal iteration layer solves the smaller systems using the
quantum algorithms. We use the HHL \cite{Harrow} and variational quantum linear solver (VQLS) \cite{Prieto} quantum algorithms for the hybrid scheme and present the theoretical
time complexity analysis for the hybrid quantum-classical scheme with the HHL and VQLS algorithms.  Two computing platforms, i.e, the statevector quantum simulator and the real quantum computer,   are considered for executing the quantum algorithms. Numerical results are presented to investigate the accuracy and efficiency of the hybrid quantum-classical scheme for solving EFIE.  Note that, a preliminary conference version of this work is presented in  \cite{Chen2025Investigation} while this work presents more implementation details and more comprehensive theoretical analyses.

\section{Fundamentals}

\subsection{EFIE}
\label{sec: EFIE}
Let $S$ denote the surface of a PEC scatterer residing in the background medium with permittivity $\varepsilon$ and permeability $\mu$. A plane wave  with the electric field $\mathbf{E}^{\mathrm{inc}}(\mathbf{r})$ is incident onto the scatterer. Upon the excitation, the current $\mathbf{J}(\mathbf{r})$ is introduced on $S$, which generates the scattered electric field $\mathbf{E}^{\mathrm{sca}}(\mathbf{r})$   in the background that can be expressed as
\begin{align}
\nonumber   \mathbf{E}^{\mathrm{sca}}(\mathbf{r})=& -j \omega \mu \int_S G(R) \mathbf{J}(\mathbf{r}') d \mathbf{r}' + \frac{1}{j \omega \varepsilon} \int_S \nabla G(R) \nabla' \cdot \mathbf{J}(\mathbf{r}') d \mathbf{r}'
\end{align}
where $\mathbf{r}$ and $\mathbf{r}'$ are the observation and source  points, $R=|\mathbf{r}-\mathbf{r}'|$, $G(R)=e^{-jkR}/(4 \pi R)$ is the time-harmonic Green function, and $k=\omega \sqrt{\mu \varepsilon}$ is the wavenumber. According to the boundary condition of the electric field, the formulation of EFIE is written as
\begin{align}
\hat{\mathbf{n}}(\mathbf{r}) \times [\mathbf{E}^{\mathrm{inc}}(\mathbf{r})+\mathbf{E}^{\mathrm{sca}}(\mathbf{r}) ]=0 , \ \ \mathbf{r} \in S.
\end{align}
Here, $\hat{\mathbf{n}}(\mathbf{r})$ is the outward normal vector at $\mathbf{r}$.

The MoM is commonly used to numerically solve (1) for $\mathbf{J}(\mathbf{r})$ as briefly introduced below. First, $S$ is meshed using flat triangular patches   resulting in $N_{\mathrm{p}}$ patches, $N_{\mathrm{n}}$ nodes, and $N_{\mathrm{e}}$ inner edges. Then, Rao-Wilton-Glisson (RWG) basis functions $\mathbf{f}_b (\mathbf{r})$, $b=1,2,...,N_{\mathrm{e}}$  \cite{Rao} are used to discretize $\mathbf{J}(\mathbf{r})$ as
\begin{align}
    \mathbf{J}(\mathbf{r})=\sum_{b=1}^{N_{\mathrm{e}}} I_b \mathbf{f}_b (\mathbf{r})
\end{align}
where  $I_b$ is the $b$th current expansion coefficient. Next, inserting (2) into (1) and applying Galerkin testing  for the resulting equation with the RWG testing functions $\mathbf{f}_a (\mathbf{r})$, $a=1,2,...,N_{\mathrm{e}}$ yield the matrix equation system of EFIE as
\begin{align}
    \mathbf{Z} \mathbf{I}=\mathbf{V}.
\end{align}
Here,  $\mathbf{I}$ is the unknown coefficient vector with the dimension of $N_{\mathrm{e}} \times 1$, $\mathbf{V}$ is the excitation vector with the dimension of $N_{\mathrm{e}} \times 1$, and  $\mathbf{Z}$ is the complex-valued, dense, and symmetric impedance matrix with the dimension of $N_{\mathrm{e}} \times N_{\mathrm{e}}$. The detailed entries of $\mathbf{I}$, $\mathbf{V}$, and $\mathbf{Z}$ can be found in \cite{Rao}. Finally, we can use different linear solvers to solve (3) for $\mathbf{I}$.  

To facilitate the use of real-valued linear solvers, we can rewrite (3) in another real-valued form as 
\begin{align}
    \mathbf{A}\mathbf{x}=\mathbf{b}.
\end{align}
Here,
\begin{align}
    \mathbf{A}=&
    \begin{bmatrix}
\mathbf{Z}_{\mathrm{real}} & \mathbf{Z}_{\mathrm{imag}} \\
\mathbf{Z}_{\mathrm{imag}} & -\mathbf{Z}_{\mathrm{real}} 
\end{bmatrix}\\
\mathbf{x}=&
\begin{bmatrix}
\mathbf{I}_{\mathrm{real}}  \\
-\mathbf{I}_{\mathrm{imag}}  
\end{bmatrix}\\
\mathbf{b}=&
\begin{bmatrix}
\mathbf{V}_{\mathrm{real}}  \\
\mathbf{V}_{\mathrm{imag}}  
\end{bmatrix}
\end{align}
where
$\mathbf{Z}_{\mathrm{real}}$, $\mathbf{Z}_{\mathrm{imag}}$, $\mathbf{I}_{\mathrm{real}}$, $\mathbf{I}_{\mathrm{imag}}$, and $\mathbf{V}_{\mathrm{real}}$, $\mathbf{V}_{\mathrm{imag}}$ are real and imaginary parts of  $\mathbf{Z}$, $\mathbf{I}$, and $\mathbf{V}$, respectively.
The dimensions of the real-valued, dense, and symmetric matrix $\mathbf{A}$, the solution vector $\mathbf{x}$, and the vector $\mathbf{b}$ are $N \times N$, $N \times 1$, and $N \times 1$, respectively, where $N=2N_{\mathrm{e}}$ is the number of unknowns of (4).

\subsection{Quantum Algorithms}
\label{sec:quantum}
Several quantum algorithms have been proposed for the solution of the linear equation system $\mathbf{A}\mathbf{x}=\mathbf{b}$. They usually first obtain the quantum state $|x \rangle$ of $\mathbf{x}$ by executing quantum circuits and then extract the classical solution vector $\mathbf{x}$ by ``measuring'' $|x \rangle$ as $|x \rangle$ has the superposition and entanglement properties before the ``measurement''. For example, the quantum state $|\psi \rangle$ of a $n$-qubit system  can be expressed in the superposition of $2^n$ basis quantum states $|i \rangle$, $i=0,1,...,2^n-1$ as
\begin{align}
    |\psi \rangle= \sum_{i=0}^{2^n-1}  a_i |i \rangle.
\end{align}
Here, $|i \rangle$ can be expressed using a unit vector with $2^n$ entries and $\sum_{i=0}^{2^n-1} |a_i|^2=1$. After the measurement, $|\psi \rangle$ collapses to the state $|i \rangle$ with the probability of $|a_i|^2$. 

Two representative quantum algorithms, the HHL and VQLS algorithms,  for solving the linear system $\mathbf{A} \mathbf{x}=\mathbf{b}$, are briefly introduced as follows.

\subsubsection{The HHL Algorithm}
\label{sec:HHL}
The HHL algorithm \cite{Harrow} is a ``direct" method that aims to directly obtain the quantum state as $|x\rangle=\mathbf{A}^{-1} |b\rangle$ via a single execution of the quantum circuit. This  algorithm requires $\mathbf{A}$ to be a Hermitian matrix, which is met by (5). 

The HHL algorithm requires 1, $m$, and $n= \lceil \mathrm{log}_2 N \rceil $ qubits for storaging  the ancilla, clock, and input-output registers, respectively, where $\lceil  \cdot \rceil $ is the ceiling function. It executes the quantum phase estimation (QPE), Pauli-$\mathbb{Y}$ rotation, and inverse QPE steps sequentially. Once the output quantum state of the ancilla register is $|1 \rangle$ after measuring the ancilla qubit, the output state of the input-output register is  $|x\rangle$.

The HHL algorithm solves the linear equation systems with the time complexity of 
\begin{align}
C_{\mathrm{single}}^{\mathrm{HHL}} \sim  \mathcal{O} \left( \frac{\kappa^2 \mathrm{log} (N)}{\xi_{\mathrm{HHL}}}     \right )   
\end{align}
where $\kappa$ denotes the condition number of $\mathbf{A}$, and $\xi_{\mathrm{HHL}}$ is the solution precision of the HHL algorithm.

\subsubsection{The VQLS Algorithm}
\label{sec:VQLS}
The VQLS algorithm \cite{Prieto} is an ``iterative" method that aims to obtain the quantum state $|x\rangle$ through iterative training of a parameterized quantum circuit. 

The parameterized quantum circuit is typically consist of simple quantum gates with repetitive structures, e.g., the most commonly used $\mathbb{R}_{\mathrm{Y}}(\Theta)$ and $\mathbb{CNOT}$ gates, where $\Theta$ denotes the circuit parameter. Once the optimal $\Theta$ is achieved by minimizing the cost function derived from the VQLS theory, the quantum state obtained from the parameterized quantum circuit can be proven to be exactly $|x \rangle$. The VQLS algorithm requires $n= \lceil \mathrm{log}_2 N \rceil $ qubits for storage.

The VQLS algorithm solves the linear equation systems with the time complexity of 
\begin{align}
C_{\mathrm{single}}^{\mathrm{VQLS}} \sim \mathcal{O}\left( \kappa \mathrm{polylog}(N) \mathrm{log} \left( \frac{1}{\xi_{\mathrm{VQLS}}}  \right) \right)   
\end{align}
where $\xi_{\mathrm{VQLS}}$ is the threshold of the cost function of the VQLS algorithm.

\section{Hybrid Quantum-Classical Scheme}
\label{sec: hybrid}
The hybrid quantum-classical scheme \cite{Ye, Chen-Z} aims to achieve high-precision computation of large-scale linear equation systems through the subspace method \cite{Saad2003} and the iterative method.

Instead of directly solving the original linear systems with the dimension of $N$ using the quantum algorithms, this hybrid scheme first designs a preconditioned linear system and then employs a double-layer iterative solution strategy, where the external iteration layer maps the original linear systems into subspace matrix equation systems with the smaller dimension of $N_{\mathrm{sub}} \ll N$ at every exterior iteration step and the internal iteration layer  solves the smaller systems using the quantum algorithms at every interior iteration step. We use the HHL and VQLS quantum algorithms for example in this work.

The detailed implementation steps of the hybrid quantum-classical scheme are presented as follows.

\noindent\rule{\columnwidth}{1.5pt}
\noindent 1) Design a  preconditioner $\mathbf{P}$ to convert the original matrix equation linear system  in (4) to the resulting preconditioned linear system $\tilde{\mathbf{A}} \mathbf{x}=\tilde{\mathbf{b}}$, where $\tilde{\mathbf{A}}=\mathbf{P}^{-1} \mathbf{A}$ and $\tilde{\mathbf{b}}=\mathbf{P}^{-1} \mathbf{b}$.

\noindent 2) Set $\mathbf{x}^{(0)}$ as the initial guess of the solution vector $\mathbf{x}$ of the preconditioned linear system.

\noindent 3) Compute the residual $ \mathbf{e}^{(0)}=  \tilde{\mathbf{b}}-\tilde{\mathbf{A}} \mathbf{x}^{(0)}$ and $\alpha^{(0)}=||\mathbf{e}^{(0)} ||_2$.

\noindent 4) Loop over the exterior iteration step $e=0,1,...,N_{\mathrm{ext}}$.

4.1) Build a subspace matrix  equation system $\mathbf{C}^{(e)} \mathbf{y}^{(e)} = \mathbf{d}^{(e)}$ 
\indent with the subspace dimension of $N_{\mathrm{sub}}$ based on the preconditioned  
\indent linear system and calculate the related coefficient 
matrix $\mathbf{U}^{(e)}$.

4.2) Set $\mathbf{y}^{(e,0)}$ as the initial guess of the solution vector $\mathbf{y}^{(e)}$ of 
\indent the subspace system.

4.3) Compute the residual $ \mathbf{f}^{(0)}= \mathbf{d}^{(e)}-\mathbf{C}^{(e)} \mathbf{y}^{(e,0)}$  and $\beta^{(0)}= \indent  || \mathbf{f}^{(0)} ||_2$.

4.4) Loop over the interior iteration step $i=0,1,...,N_{\mathrm{int}}^{e}$.

\indent \indent 4.4.1) Solve the linear equation system $\mathbf{C}^{(e)} \mathbf{z}^{(e,i)} = \mathbf{f}^{(i)}$ using 
\indent  \indent  the quantum algorithms (e.g, the HHL and VQLS algorithms 
\indent  \indent  considered in this work) for the quantum state $|z \rangle ^{(e,i)}$.

\indent \indent 4.4.2) Extract the classical solution vector $\hat{\mathbf{z}}^{(e,i)}$ from $|z \rangle ^{(e,i)}$.

\indent \indent 4.4.3) Compute $|| \mathbf{z}^{(e,i)}||_2$ based on  the principle of minimum 
\indent \indent $L_2$-norm.

\indent \indent 4.4.4)  Obtain  $\mathbf{z}^{(e,i)}= || \mathbf{z}^{(e,i)}||_2 \hat{\mathbf{z}}^{(e,i)}$.

\indent \indent 4.4.5)  Update  $\mathbf{y}^{(e,i+1)}=\mathbf{y}^{(e,i)}+\mathbf{z}^{(e,i)}$ for the next interior 
\indent \indent iteration step.

\indent \indent 4.4.6)  Update the residual  $\mathbf{f}^{(i+1)}=\mathbf{f}^{(i)}-\mathbf{C}^{(e)} \mathbf{z}^{(e,i)}$ and 
\indent \indent $\beta^{(i+1)}=|| \mathbf{f}^{(i+1)} ||_2$ for the next interior iteration step.

\indent \indent 4.4.7) Check if the convergence condition $\beta^{(i+1)} \le \xi_{\mathrm{int}}$ is 
\indent \indent satisfied, where $\xi_{\mathrm{int}}$ is the  convergence threshold of the internal 
\indent \indent iteration layer. Upon 
convergence, the solution  vector of the 
\indent \indent subspace system is obtained as $\mathbf{y}^{(e)}=\mathbf{y}^{(e,i+1)}$   and end the 
\indent \indent loop over $i$.

\indent 4.5) Update   $\mathbf{x}^{(e+1)}=\mathbf{x}^{(e)}+\mathbf{U}\mathbf{y}^{(e)}$ for the next exterior iteration 
\indent step.

\indent 4.6)  Update the residual  $ \mathbf{e}^{(e+1)}=  \tilde{\mathbf{b}}-\tilde{\mathbf{A}} \mathbf{x}^{(e+1)}$ and 
$\alpha^{(e+1)}= \indent  ||\mathbf{e}^{(e+1)} ||_2$ for the next exterior iteration step.

\indent 4.7) Check if the convergence condition
$    \alpha^{(e+1)} \le \xi_{\mathrm{ext}}$ is satisfied, 
\indent where 
$\xi_{\mathrm{ext}}$ is the 
convergence threshold of the external iteration 
\indent layer.  Upon convergence, 
the  solution vector of the preconditioned  
\indent linear system is obtained as  $\mathbf{x}=\mathbf{x}^{(e+1)}$ and end the loop over $e$.
\noindent\rule{\columnwidth}{1.5pt}

As seen from Step 4.1 to Step 4.7, the time cost of the hybrid quantum-classical scheme at each exterior iteration step  is mainly composed of two parts, the build of the subspace linear equation system at Step 4.1  with the time cost of $T_{\mathrm{sub}}^{\mathrm{build}}$ and the solution of the subspace system using the internal iterative quantum algorithms at Step 4.4 with the time cost of $T_{\mathrm{iter}}^{\mathrm{quantum}}$. Therefore, the total time cost of the hybrid quantum-classical scheme $T_{\mathrm{hybrid}}^{\mathrm{quantum}}$ can be expressed as
\begin{align}
    T_{\mathrm{hybrid}}^{\mathrm{quantum}} \approx N_{\mathrm{ext}} ( T_{\mathrm{sub}}^{\mathrm{build}} + T_{\mathrm{iter}}^{\mathrm{quantum}}  ).
\end{align}
Accordingly, we can derive the theoretical time complexity of the hybrid quantum-classical scheme with the HHL algorithm (termed as ``hybrid HHL-classical scheme'') and the hybrid scheme with the VQLS algorithm (termed as ``hybrid VQLS-classical scheme'') based on (9) and (10), respectively. Therefore, the time complexity of the hybrid HHL-classical scheme is  
\begin{align}
    C_{\mathrm{hybrid}}^{\mathrm{HHL}} \sim \mathcal{O} \left ( \frac{\tilde{\kappa} \mathrm{log}(\xi_{\mathrm{ext}})}{N_{\mathrm{sub}} \mathrm{log}(\xi_{\mathrm{int}})}  \left(N N_{\mathrm{sub}} +\frac{\bar {\kappa}_{\mathrm{sub}}^2 \mathrm{log} (N_{\mathrm{sub}})}{\xi_{\mathrm{int}}^4 \xi_{\mathrm{HHL}}}     \right ) \right ) 
\end{align}
and that of the hybrid VQLS-classical scheme is
\begin{align}
 \nonumber  & C_{\mathrm{hybrid}}^{\mathrm{VQLS}} \sim  \\
   & \mathcal{O} \left ( \frac{ \tilde {\kappa} \mathrm{log}(\xi_{\mathrm{ext}})}{N_{\mathrm{sub}} \mathrm{log}(\xi_{\mathrm{int}})}  \left(N N_{\mathrm{sub}} +\frac{\bar {\kappa}_{\mathrm{sub}} \mathrm{polylog}(N_{\mathrm{sub}})}{\xi_{\mathrm{int}}^4}  \mathrm{log} \left( \frac{1}{\xi_{\mathrm{VQLS}}} \right) \right ) \right ) 
\end{align}
where  $\tilde{\kappa}$ is the condition number of the preconditioned matrix $\tilde{\mathbf{A}}$, $\bar {\kappa}_{\mathrm{sub}}$ represents the average condition number of the subspace matrices over all the exterior iteration steps, $\frac{\tilde{\kappa} \mathrm{log}(\xi_{\mathrm{ext}})}{N_{\mathrm{sub}} \mathrm{log}(\xi_{\mathrm{int}})}$ denotes the scaling of $N_{\mathrm{ext}}$, $N N_{\mathrm{sub}}$ denotes the computational complexity of the build of the subspace linear equation system at Step 3.1, and $\frac{1}{\xi_{\mathrm{int}}^4}$ accounts for the scaling of $N_{\mathrm{int}}^{e}$.

\begin{figure}[t]
    \centering
    \includegraphics[width=1\columnwidth]{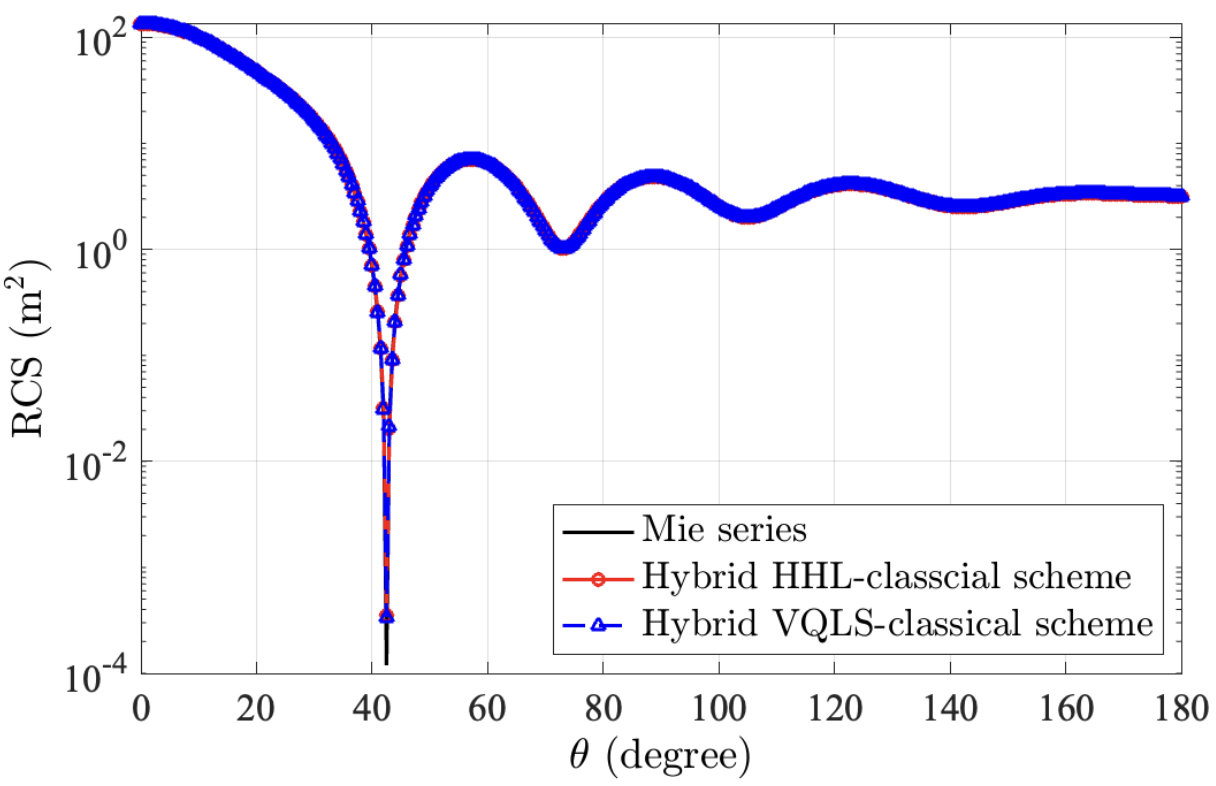}
    \caption{Comparison of the RCS results of the PEC unit sphere obtained after  using the hybrid HHL- and VQLS-classical schemes with those obtained using the Mie series analytical solution with respect to $\theta =[0^{\mathrm{o}}, 180^{\mathrm{o}}] $ at 300 MHz.}
\end{figure}

\begin{figure}[t]
    \centering
    \subfigure[]{\includegraphics[width=0.494\columnwidth]{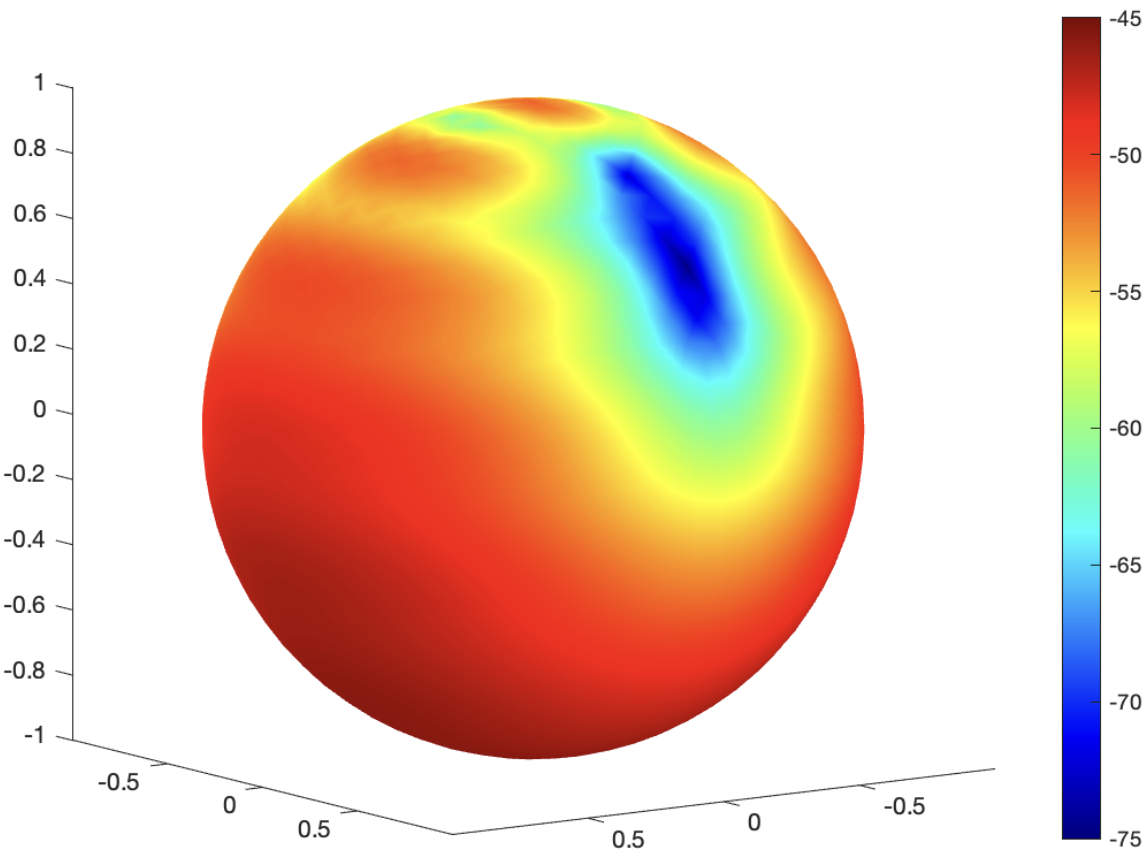}} \\
    \subfigure[]{\includegraphics[width=0.494\columnwidth]{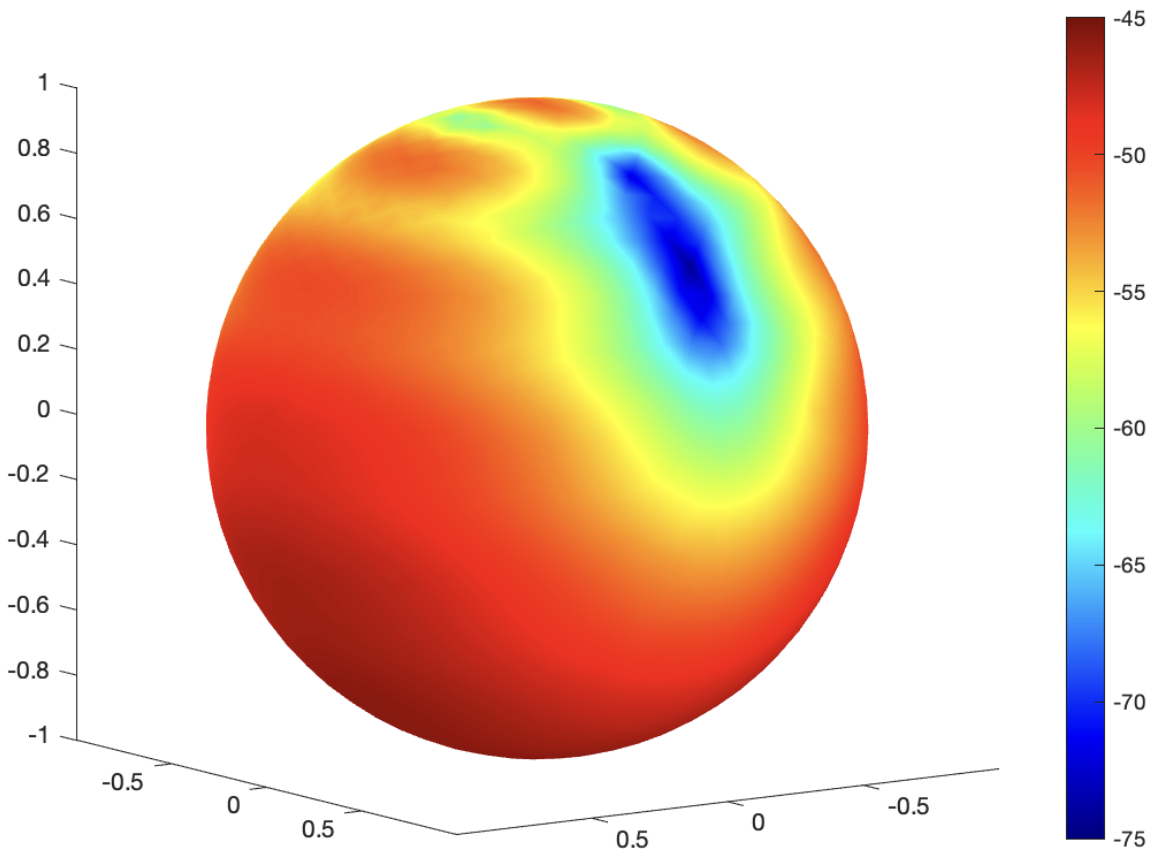}} 
    \subfigure[]{\includegraphics[width=0.494\columnwidth]{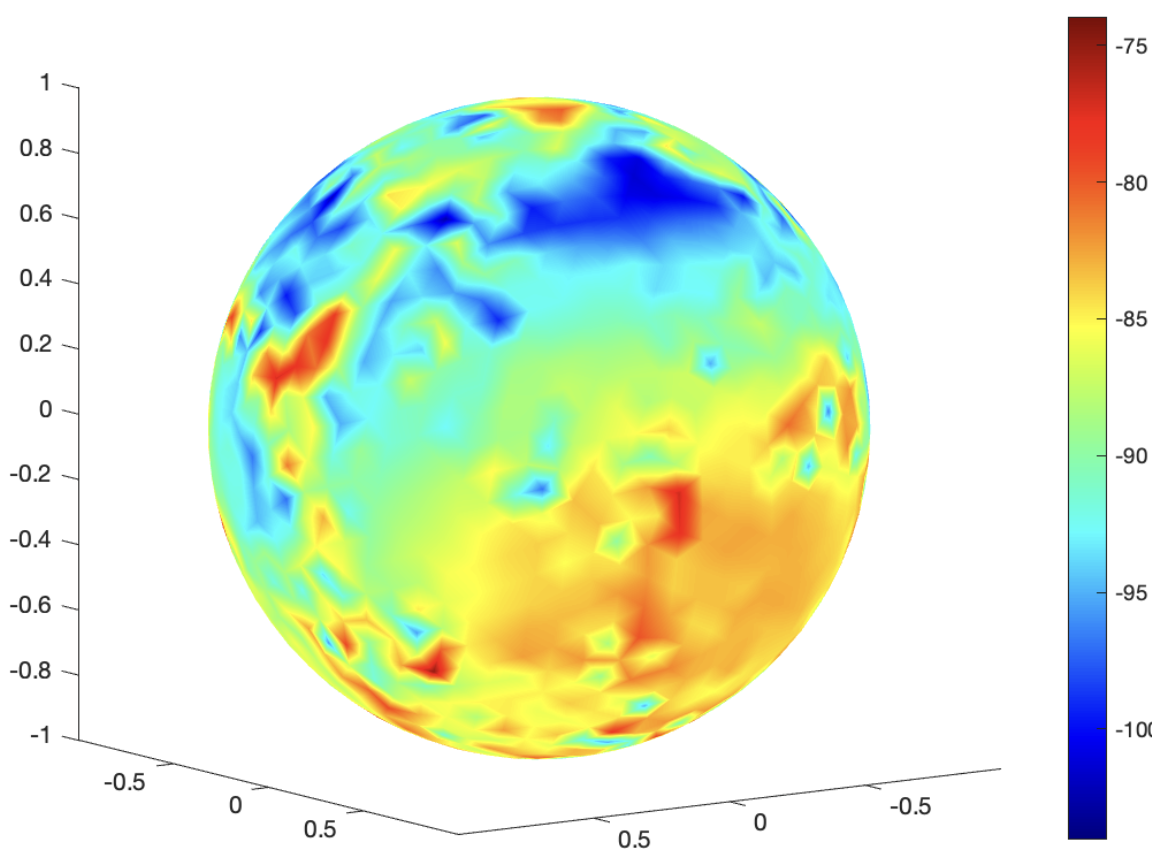}} \\
    \subfigure[]{\includegraphics[width=0.494\columnwidth]{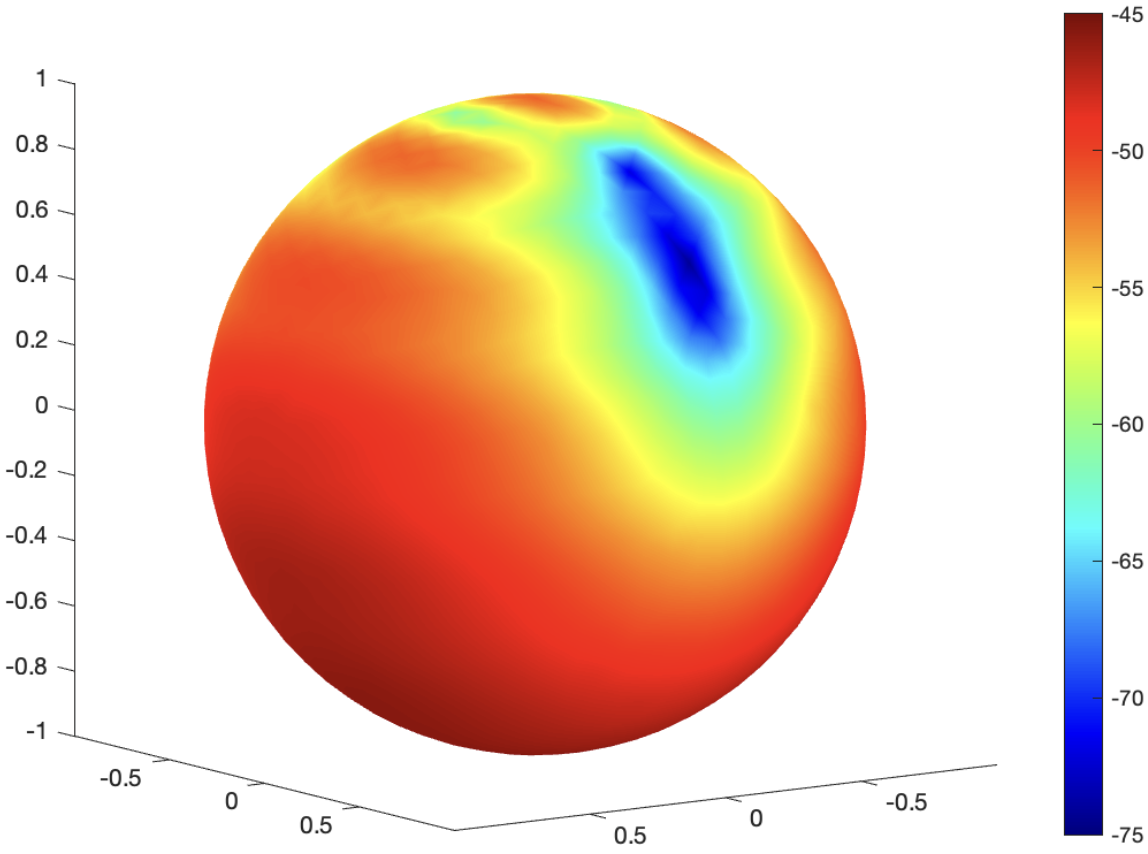}} 
    \subfigure[]{\includegraphics[width=0.494\columnwidth]{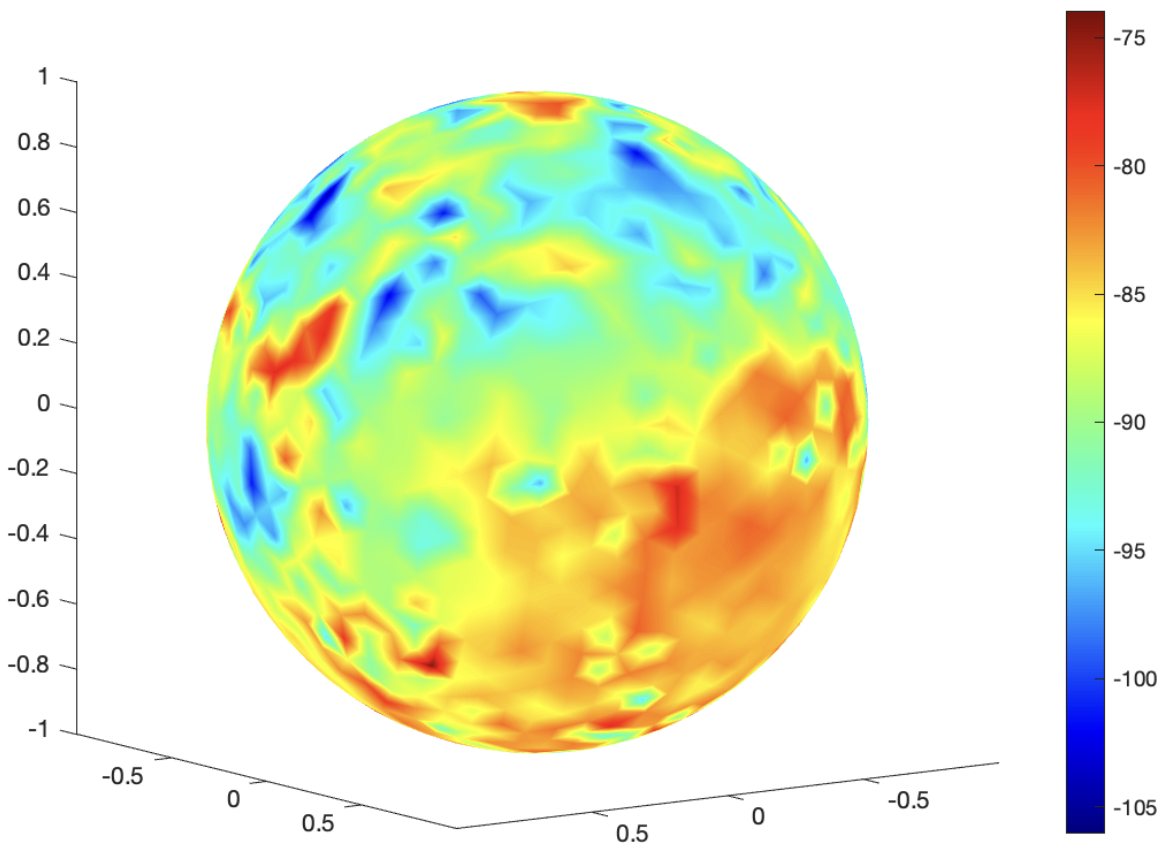}} 
    \caption{The surface current distributions of the PEC unit sphere obtained after using (a) the Mie series solution, (b) the hybrid HHL-classical scheme, (c) the difference between (a) and (b), (d) the hybrid VQLS-classical scheme, and (e) the difference between (a) and (d).}
\end{figure}

\section{Numerical Results}
\label{sec:results}
In this section,  several numerical experiments are presented to demonstrate the accuracy and efficiency of the hybrid quantum-classical scheme for solving EFIE for the analysis of electromagnetic scattering from 3D PEC objects in free space under the incident plane wave with the electric field 
$ \mathbf{E}^{\mathrm{inc}}(\mathbf{r})=   \hat{\mathbf{x}} e^{-jk z} $.  

To quantify the accuracy of the solution of EFIE, the relative error of the radar cross section (RCS) is defined as
\begin{align}
    \delta_{\mathrm{RCS}}  =\frac{|| \sigma^{\mathrm{quantum}}_{\mathrm{hybrid}}(\theta,\phi)- \sigma^{\mathrm{ref}}(\theta,\phi) ||_2}{|| \sigma^{\mathrm{ref}}(\theta,\phi) ||_2}
\label{eq52}
\end{align}
where $\theta =[0^{\mathrm{o}}, 180^{\mathrm{o}}] $ and $\phi=0^{\mathrm{o}}$ are the observation angles of the far field,  $\sigma^{\mathrm{quantum}}_{\mathrm{hybrid}}(\theta, \phi)$ represents the RCS results obtained after solving the EFIE matrix equation system using the hybrid quantum-classical scheme, and $\sigma^{\mathrm{ref}}(\theta, \phi)$ represents the RCS results obtained after using other reference methods.

In the following sections, the single-layer hardware-efficient ansatz quantum circuit \cite{Kandala} is used for the VQLS algorithm with $\xi_{\mathrm{VQLS}}=10^{-3}$.  Two computing platforms, the quantum simulator and the quantum computer, are considered for executing the quantum algorithms.

\begin{table*}
\caption{Effect of twelve cases of the parameters on the performance of the hybrid quantum-classical scheme for solving EFIE.}
\centering
\begin{tabular}{ c|c|c|c|c|c|c|c|c|c|c } 
 \hline
 \hline
 Case & Quantum algorithm &  Noise of gates  & $\tilde{\kappa}$ & $N_{\mathrm{sub}}$ & $\xi_{\mathrm{ext}}$ & $\xi_{\mathrm{int}}$   & $N_{\mathrm{qubit}}$  & $N_{\mathrm{ext}}$ & $N_{\mathrm{int}}$ & $ \delta_{\mathrm{RCS}}  $ \\ 
 \hline
 1 & HHL & No  & 75.27 & 32 & $10^{-3}$ & $10^{-3}$   & 12 & 5  & 10 & 0.0061 \\ 
 \hline
 2 & HHL & No  & 75.27 & 32 & $10^{-3}$ & $10^{-3}$   & 16 & 5  & 9 & 0.0059 \\ 
 \hline
 3 & HHL & No  & 5.23 & 32 & $10^{-3}$ & $10^{-3}$   & 21 & 1  & 1 & 0.0049 \\ 
 \hline
 4 & HHL & No  & 5.23 & 32 & $10^{-3}$ & $10^{-4}$   & 21 & 1  & 1 & 0.0049 \\ 
 \hline
 5 & HHL & No  & 5.23 & 32 & $10^{-2}$ & $10^{-3}$   & 21 & 1  & 1 & 0.0049 \\ 
 \hline
 6 & HHL & No  & 5.23 & 4 & $10^{-3}$ & $10^{-3}$   & 8 & 1  & 3 & 0.0055 \\ 
 \hline
 7 & VQLS & No  & 75.27 & 32 & $10^{-3}$ & $10^{-3}$  &  5 & 40 & 192432  & 0.0058\\ 
 \hline
 8 & VQLS & No  & 5.23 & 32 & $10^{-3}$ & $10^{-3}$  &  5 & 1 & 287  & 0.0047\\ 
 \hline
 9 & VQLS & No  & 5.23 & 32 & $10^{-3}$ & $10^{-4}$  &  5 & 1 & 1673  & 0.0052\\ 
 \hline
 10 & VQLS & No  & 5.23 & 32 & $10^{-2}$ & $10^{-3}$  &  5 & 1 & 4  & 0.0054\\ 
 \hline
 11 & VQLS & No  & 5.23 & 4 & $10^{-3}$ & $10^{-3}$  &  2 & 2 & 4  & 0.0047\\ 
 \hline
 12 & VQLS & Yes  & 5.23 & 32 & $10^{-3}$ & $10^{-3}$  &  5 & 1 & 99  & 0.0058\\ 
 \hline
 \hline
\end{tabular}
\end{table*}

\begin{table}
\caption{Time cost of the hybrid VQLS-classical scheme for solving the EFIE matrix equation system $\mathbf{A} \mathbf{x}=\mathbf{b}$ with the preconditioner $\mathbf{P}=\mathbf{I}$.}
\centering
\begin{tabular}{ c|c|c|c|c|c|c|c } 
 \hline
 \hline
 $r$ (m)   & $N_{\mathrm{p}}$ & $N_{\mathrm{n}}$ & $N_{\mathrm{e}}$  & $N$  & $\tilde{\kappa}$ & $\bar{\kappa}_{\mathrm{sub}}   $ & Time (s)  \\ 
 \hline
  0.25 & 208  & 106 & 312 & 624 & 99.82 & 26.53  & 32.09   \\ 
 \hline
  0.25 & 322  & 163 & 483 & 966 & 129.12 & 36.09 & 67.70  \\ 
 \hline
  0.5 & 808  & 406 & 1212 & 2424 & 108.17 & 39.10  &  168.85 \\ 
 \hline
  0.5 & 1662  & 833 & 2493 & 4986 & 219.08 & 28.72 & 113.72  \\ 
 \hline
  1 & 3390  & 1697 & 5085 & 10170 & 109.79 &  35.63 & 497.78  \\ 
 \hline
 \hline
\end{tabular}
\end{table}

\begin{figure}[t]
    \centering
    \includegraphics[width=1\columnwidth]{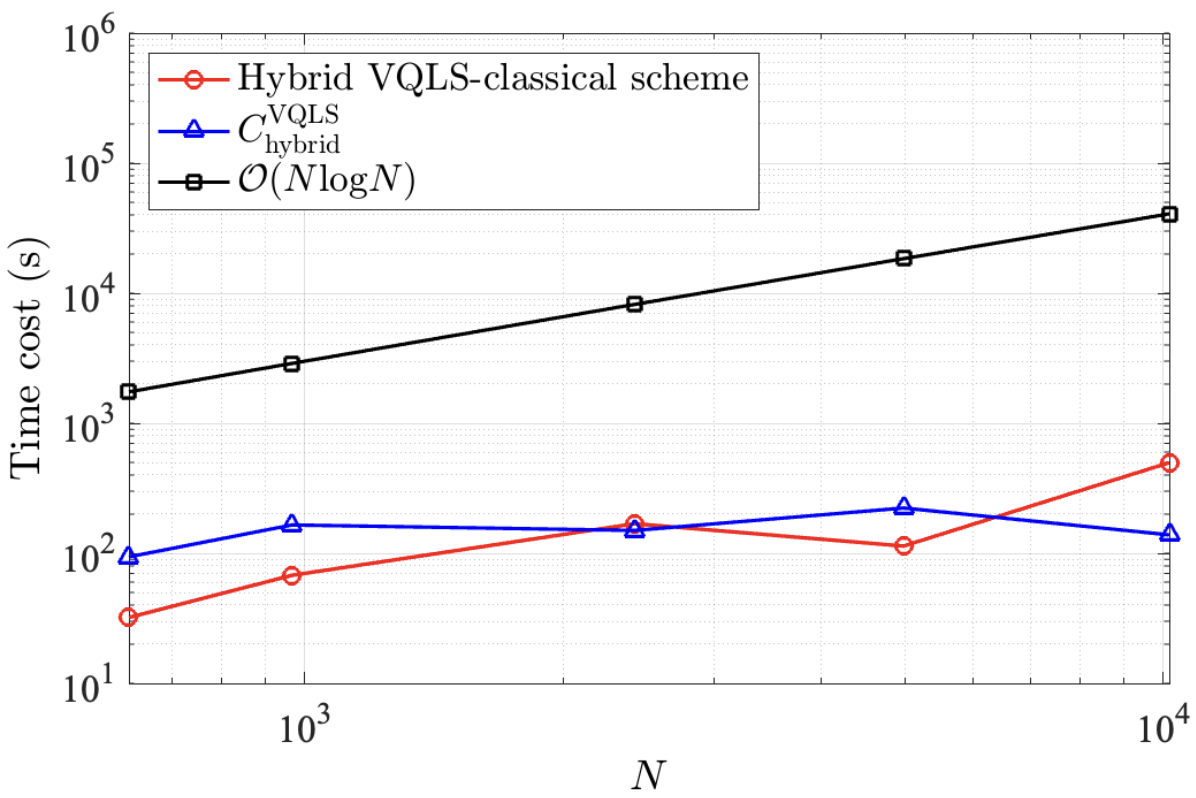}
    \caption{Comparison of the time cost of the hybrid VQLS-classical scheme with its theoretical computational complexity presented in (13) with respect to the number of unknowns.}
\end{figure}

\subsection{Quantum Simulator}
\label{sec:simulator}
The first computing platform is a virtual statevector quantum simulator developed based on the \textit{QPanda} library \cite{QPanda}, which is  executed  on classical computers.  The 3D  scatterers considered in this section are the PEC spheres under the incidence of the plane wave at $300 \ \mathrm{MHz}$. The Mie series analytical solution \cite{Harrington2001} is used as the reference method in (14).

First, the PEC unit sphere (with the radius of $r=1 \ \mathrm{m}$) is  meshed with $N_{\mathrm{p}}=2982$, $N_{\mathrm{n}}=1493$, and $N_{\mathrm{e}}=4473$. The HHL and VQLS quantum algorithms are performed on an Intel Core i7-13700 CPU at 2.1 GHz.  No noise of the quantum gates is considered. The incomplete LU factorization with the drop tolerance $\xi_{\mathrm{ILU}}=10^{-3}$ \cite{Saad1996}  is used for the design of the preconditioner at Step 1 in Section III and the condition number of the preconditioned matrix $\tilde{\mathbf{A}}$ is 5.23. The subspace dimension is $N_{\mathrm{sub}}=32$. The convergence thresholds of the external and internal
iteration layers of the hybrid quantum-classical scheme are $\xi_{\mathrm{ext}}=10^{-3}$ and $\xi_{\mathrm{int}}=10^{-3}$, respectively. The hybrid HHL-classical scheme and the hybrid VQLS-classical scheme require the number of qubits $N_{\mathrm{qubit}}=21$ and $N_{\mathrm{qubit}}=5$ for storage, respectively. Figure 1 compares the RCS results of the PEC unit sphere obtained after using the hybrid HHL- and VQLS-classical schemes with those obtained using the Mie series analytical solution with respect to $\theta =[0^{\mathrm{o}}, 180^{\mathrm{o}}] $ at $300$ MHz.  As seen from Fig. 1, both results obtained after using the hybrid HHL- and VQLS-classical schemes match very well with the analytical Mie series result. The RCS relative error values of the hybrid HHL-classical scheme and the hybrid VQLS-classical scheme are $\delta_{\mathrm{RCS}}=0.0049$ and $\delta_{\mathrm{RCS}}=0.0047$, respectively. In addition, Figure 2 plots the surface current distributions of the PEC unit sphere obtained after using (a) the Mie series solution, (b) the hybrid HHL-classical scheme, (c) the difference between (a) and (b), (d) the hybrid VQLS-classical scheme, and (e) the difference between (a) and (d). Clearly, the hybrid quantum-classical scheme with both quantum algorithms  obtain the accurate  current distributions on the sphere surface after solving the EFIE linear system.

To further investigate the effect of different parameters on the performance of the hybrid quantum-classical scheme for solving EFIE, twelve cases of the parameters (named as ``Case-1", ``Case-2", ...,  ``Case-12") are considered as presented in Table I, where $N_{\mathrm{int}}= \sum_{e=0}^{N_{\mathrm{ext}}} (N_{\mathrm{int}}^{e}+1)-1 $ represents the maximum number of the  ``global'' interior iteration steps (i.e., $0,1,...,N_{\mathrm{int}}$) of the internal iteration layer.  As seen from Table I, all the cases of the hybrid quantum-classical scheme provide the  good solution accuracy as all the $\delta_{\mathrm{RCS}}$ values are at the level of $10^{-3}$. Besides, by comparing Case-8 (no noise of the gates) with Case-12 (the noise error of the $\mathbb{R}_{\mathrm{Y}}(\Theta)$ and $\mathbb{CNOT}$ gates is set to be $0.2$) in Table I, it can be found that 
the introduction of the noise of gates causes a slight decrease in accuracy of the solution of the hybrid VQLS-classical scheme. Furthermore, it can be found that smaller condition number of the linear system yields less number of $N_{\mathrm{ext}}$ and $N_{\mathrm{int}}$ by comparing Case-1 and Case-7 (set the preconditioner $\mathbf{P}$ to be the identity matrix $\mathbf{I}$ at Step 1 in Section III) with Case-3 and Case-8 (use the incomplete LU factorization with the drop tolerance $\xi_{\mathrm{ILU}}=10^{-3}$  \cite{Saad1996} for the design of the preconditioner at Step 1 in Section III) for the hybrid HHL- and VQLS-classical schemes in Table I, respectively.   Moreover, Table I indicates that $N_{\mathrm{qubit}}$ increases as $N_{\mathrm{sub}}$ becomes larger by comparing Case-3 with Case-6 for the hybrid HHL-classical scheme and comparing Case-8 with Case-11 for the hybrid  VQLS-classical scheme. It should be noted that, as a comparison, the original VQLS algorithm  in Section II-B requires  $14$ qubits for  solving the same linear system, which is more than the number of qubits required by Case-8 of the hybrid VQLS-classical scheme. Additionally, when the solution precision of the HHL algorithm is improved (i.e., $\xi_{\mathrm{HHL}}$ decreases and $N_{\mathrm{qubit}}$ increases),  the solution of the linear system using the hybrid HHL-classical scheme is more accurate by comparing Case-1 and Case-2 in Table I. In addition, as seen from Table I, $N_{\mathrm{ext}}$ is more resistant to the variation of $\xi_{\mathrm{ext}}$ and $\xi_{\mathrm{int}}$ for the hybrid HHL- and VQLS-classical schemes while $N_{\mathrm{int}}$ for the hybrid VQLS-classical scheme is more prone to the variation of $\xi_{\mathrm{ext}}$ and $\xi_{\mathrm{int}}$ than that for the HHL-classical scheme.

\begin{figure}[t]
    \centering
    \includegraphics[width=0.3\columnwidth]{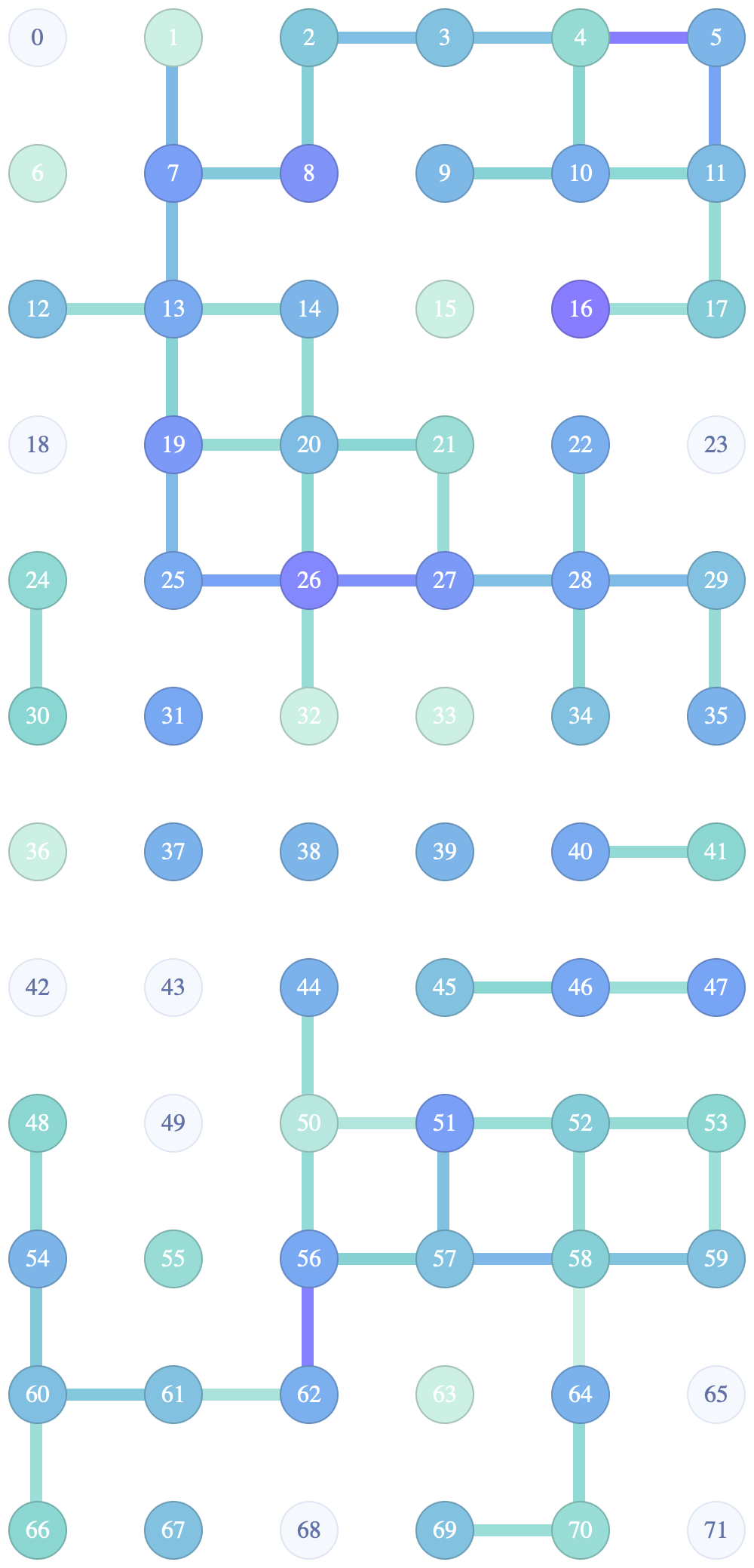}
    \caption{The topology structure of the 72-qubit quantum chip.}
    \label{fig:placeholder}
\end{figure}

\begin{figure}[t]
    \centering
    \includegraphics[width=1\columnwidth]{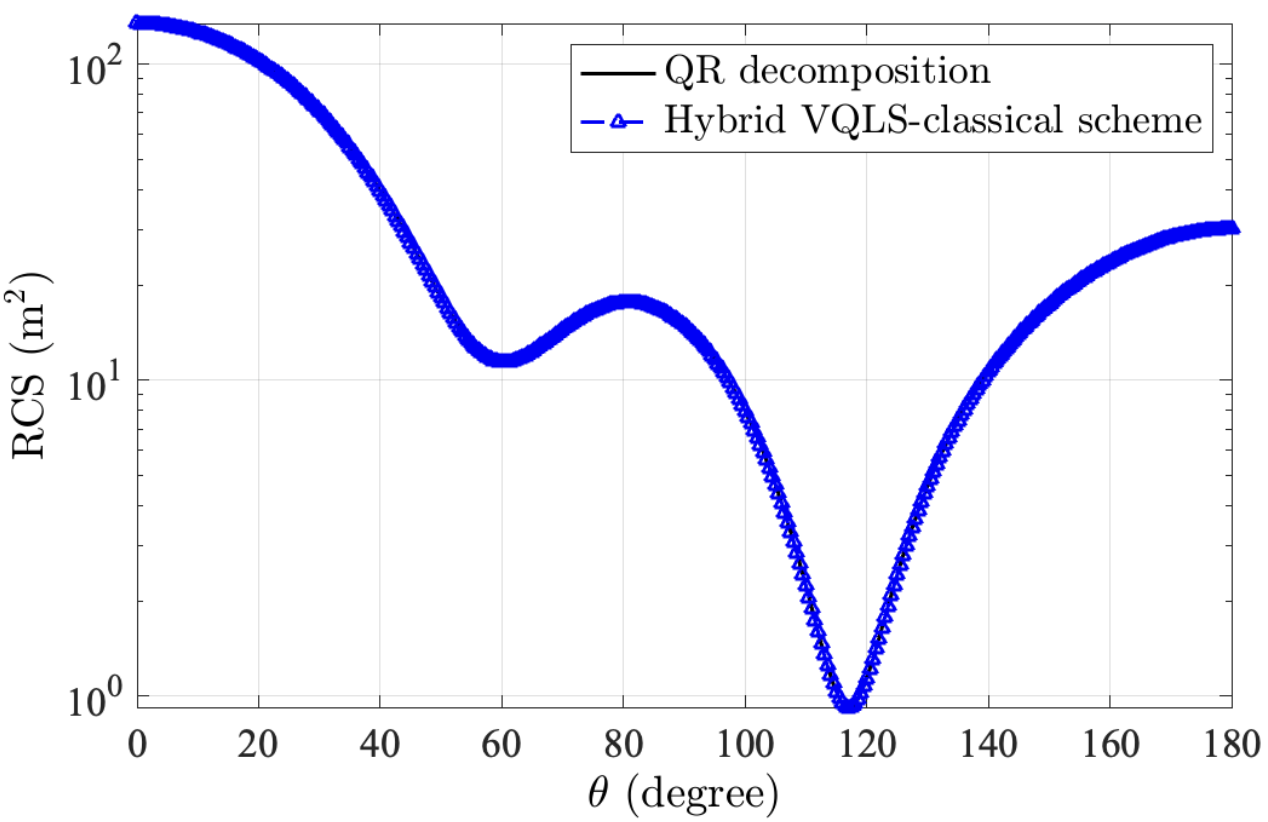}
    \caption{Comparison of the RCS results of the PEC flower-shaped scatterer obtained after  using the hybrid VQLS-classical scheme with those obtained using the QR decomposition method with respect to $\theta =[0^{\mathrm{o}}, 180^{\mathrm{o}}] $ at 69.89 MHz.}
\end{figure}

\begin{figure}[t]
    \centering
    \subfigure[]{\includegraphics[width=0.494\columnwidth]{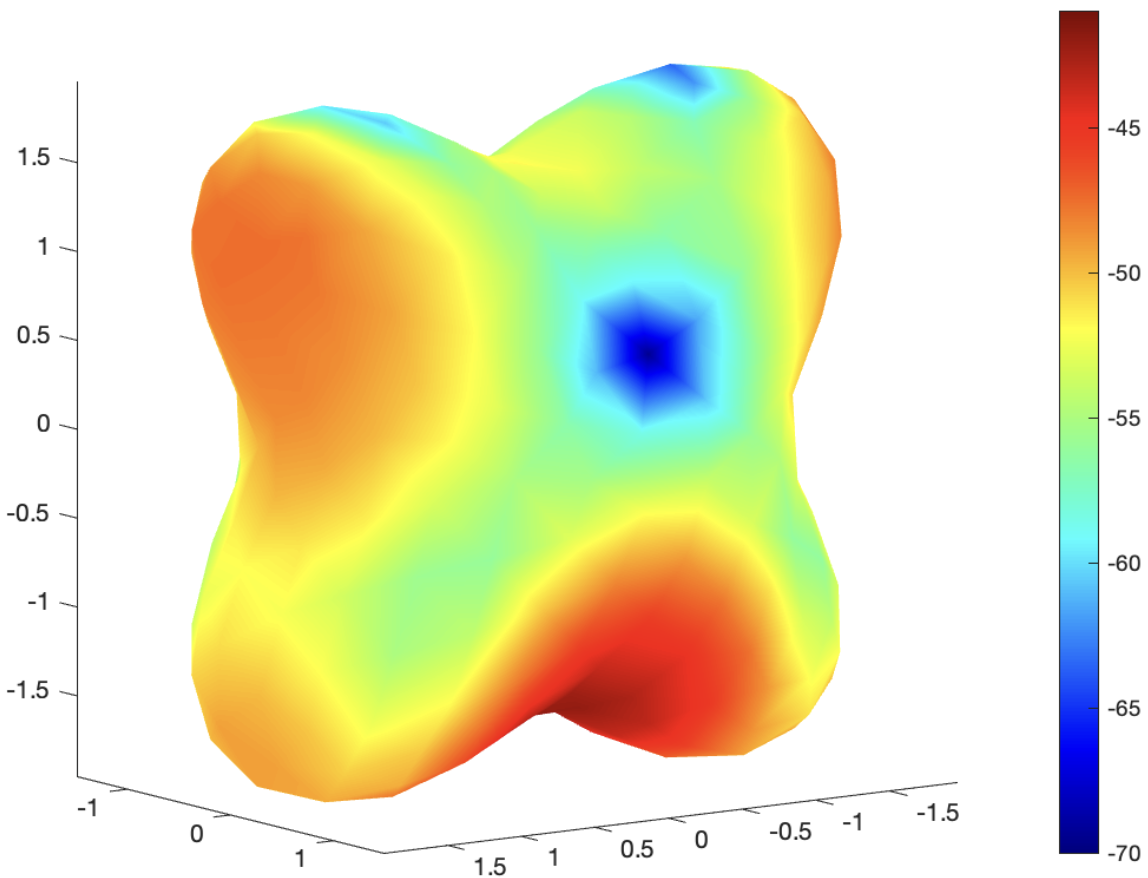}} 
    \subfigure[]{\includegraphics[width=0.494\columnwidth]{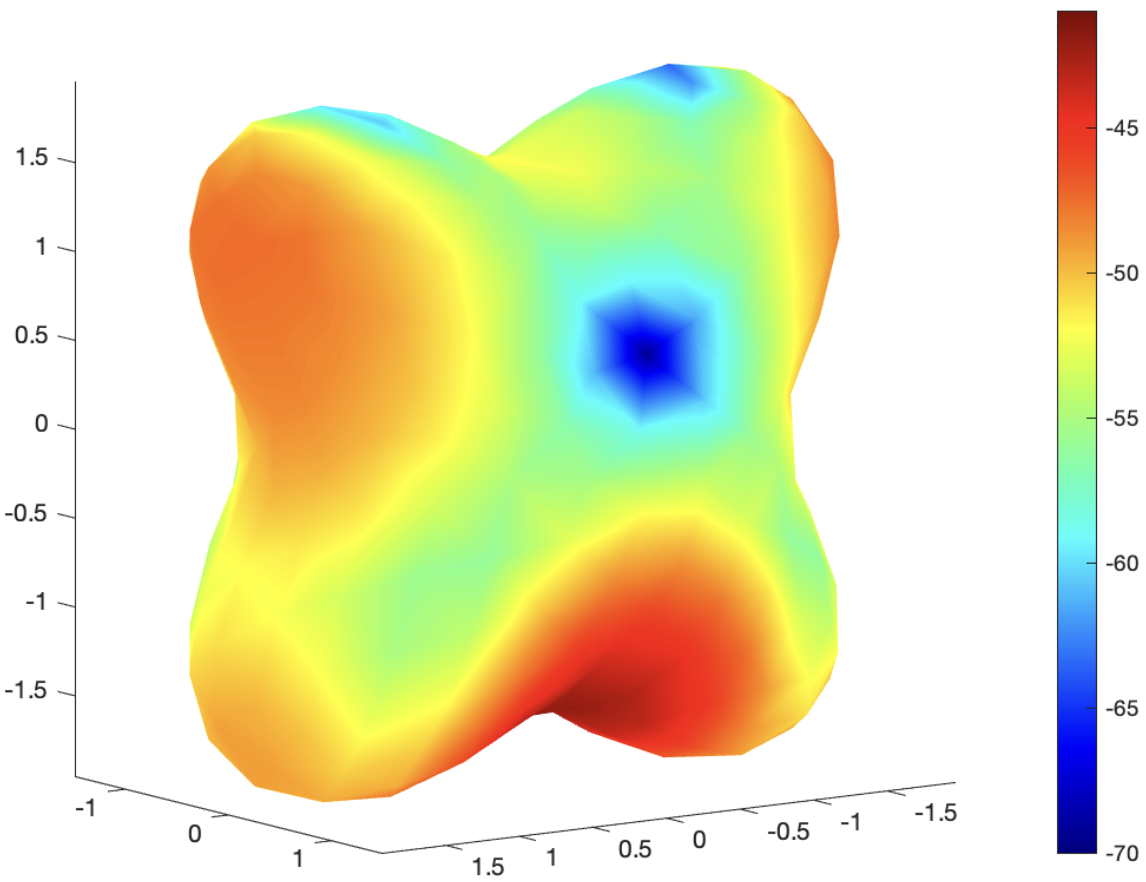}} 
    \subfigure[]{\includegraphics[width=0.494\columnwidth]{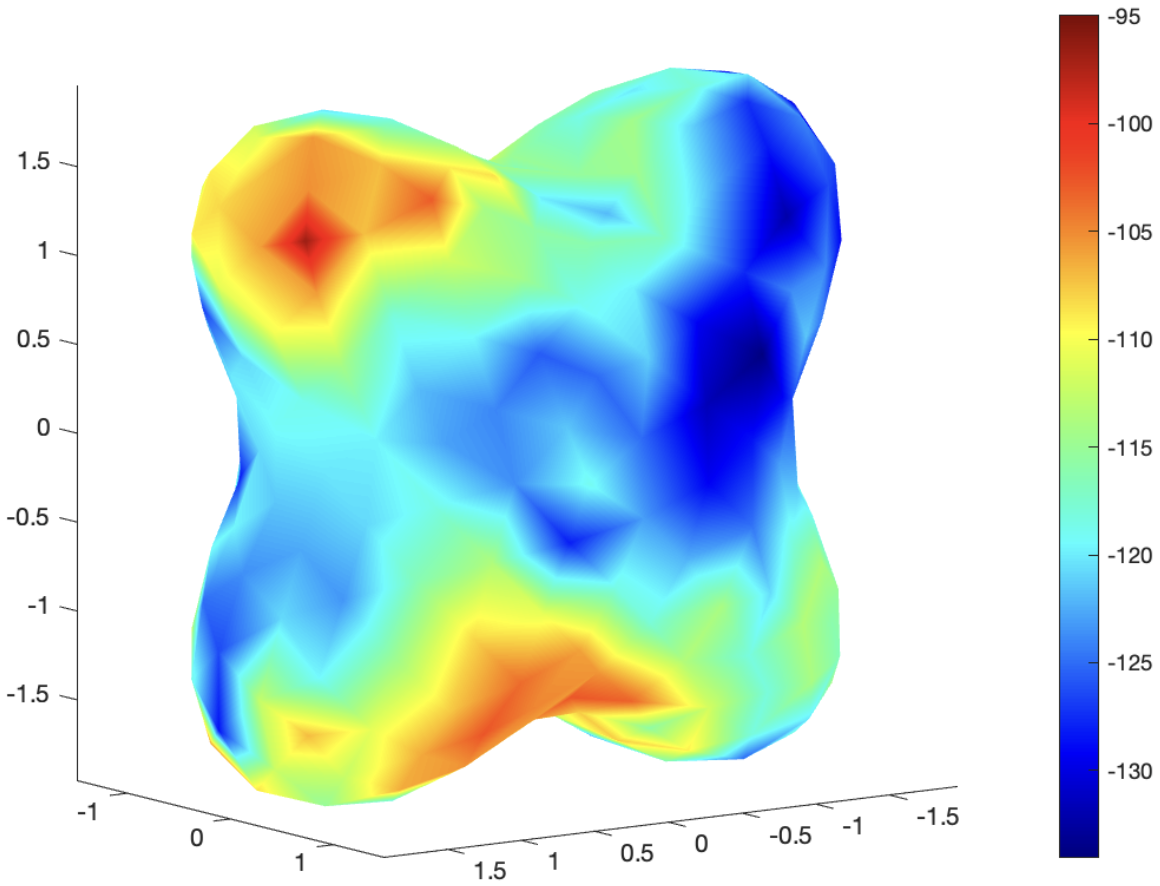}} 
    \caption{The surface current distributions of the PEC flower-shaped scatterer  obtained after using (a) the hybrid VQLS-classical scheme, (b) the QR decomposition method,  and (c) the difference between (a) and (b).}
\end{figure}

Next, to investigate the computational efficiency of the hybrid quantum-classical scheme, five discretization sets of the PEC spheres are considered and  the time cost of the hybrid VQLS-classical scheme for solving the EFIE matrix equation system $\mathbf{A} \mathbf{x}=\mathbf{b}$ with the preconditioner $\mathbf{P}=\mathbf{I}$ is provided as shown in Table II. The VQLS quantum algorithm is executed on an Intel Core i5-10500 CPU at 3.1 GHz. Figure 3 compares the time cost of the hybrid VQLS-classical scheme with its theoretical computational complexity presented in (13) with respect to the number of unknowns. This figure shows that the curve of the practical time cost of the hybrid VQLS-classical scheme aligns well with that of its theoretical computational complexity in (13), which demonstrates the accuracy of our time complexity analysis in Section III. Additionally, Figure 3 shows that the time complexity curve of the hybrid VQLS-classical scheme performs better than that of the conventional fast solvers  in classical computing (in order of $\mathcal{O} (N \mathrm{log} N)$), which demonstrates the potential advantage of the hybrid quantum-classical scheme over the conventional classical solvers for solving large-scale electromagentic problems.

\subsection{Quantum Computer}
\label{sec:NISQ}
The other computing platform is a  superconducting quantum computer \textit{OriginQ Wukong} with the intermediate-scale noise  \cite{qcloud}. The calibration parameters of this quantum hardware are 72 qubits, the average relaxation time  $T_1=18.87  $ µs,
the average dephasing time $T_2=5.494$ µs, the average 1-qubit gate fidelity $0.9972$, the average  2-qubit gate fidelity $0.9629$, and the average readout fidelity $0.8736$. The topology structure of the 72-qubit quantum chip with the 4-regular topology is presented in Fig. 4.

The 3D scatterer considered in this
section is the PEC flower-shaped scatterer (see \cite{Wildman} for its model definition) meshed with $N_{\mathrm{p}}=588$, $N_{\mathrm{n}}=296$, and $N_{\mathrm{e}}=882$ under the incident of the plane wave at $69.89 \ \mathrm{MHz}$. The hybrid quantum-classical scheme using the QR decomposition for the solution of the linear equation system at Step 4.4.1 in Section III is considered as the reference method in (14).

The VQLS quantum algorithm  is executed on the superconducting quantum computer. The incomplete LU factorization with the drop tolerance $\xi_{\mathrm{ILU}}=10^{-3}$ \cite{Saad1996} is used for the design the  preconditioner at Step 1 in Section III. The subspace dimension is $N_{\mathrm{sub}}=16$. The convergence thresholds of the external and internal iteration layers of the hybrid  VQLS-classical scheme are $\xi_{\mathrm{ext}}=10^{-3}$ and $\xi_{\mathrm{int}}=10^{-3}$, respectively. The hybrid VQLS-classical scheme requires  four  qubits for storage (i.e., $N_{\mathrm{qubit}}=4$). Specifically, the $20 \mathrm{th}$, $26 \mathrm{th}$, $27 \mathrm{th}$, and $28 \mathrm{th}$ physical qubits  in Fig. 4  are selected for the execution of the VQLS algorithm on the real quantum hardware.
The CZ gate fidelity between the $20 \mathrm{th}$ and $26 \mathrm{th}$ qubits is $0.9648$, the CZ gate fidelity between the $26 \mathrm{th}$ and $27 \mathrm{th}$ qubits is $0.9855$, and the CZ gate fidelity between the $27 \mathrm{th}$ and $28 \mathrm{th}$ qubits is $0.9739$. The maximum numbers of the global exterior and interior iteration steps of the external and internal iteration layers are $N_{\mathrm{ext}}=84$ and $N_{\mathrm{int}}=189$, respectively. 

Figure 5 compares the RCS results of the PEC flower-shaped scatterer obtained after  using the hybrid VQLS-classical scheme with those obtained using the  QR decomposition method with respect to $\theta =[0^{\mathrm{o}}, 180^{\mathrm{o}}] $ at $69.89 \ \mathrm{MHz}$.  This figure clearly shows that the RCS result obtained after using the hybrid VQLS-classical scheme matches very nice with the QR decomposition result.  The relative error of RCS of the hybrid VQLS-classical scheme is $\delta_{\mathrm{RCS}}=1.502 \times 10^{-5}$. In addition, Figure 6 plots the surface current distributions of the PEC flower-shaped scatterer  obtained after using (a) the hybrid VQLS-classical scheme, (b) the QR decomposition method,  and (c) the difference between (a) and (b). These figures clearly show that the  current distribution on the surface of the flower-shaped scatterer obtained after using the hybrid VQLS-classical scheme for solving the EFIE linear system is accurate.

In addition, the time cost of the  hybrid VQLS-classical scheme for solving the EFIE matrix equation system is $12028.1 \ \mathrm{s}$ while the time cost  for executing the VQLS algorithm on the real quantum chip is $9074.62 \ \mathrm{s}$. It needs to be mentioned that, the time cost of the quantum algorithm performed on the real quantum computer is usually much larger than that performed on the statevector quantum simulator. This is due to the fact that the statevector quantum simulator can directly calculate the exact expectation value using linear algebra operations while the real quantum hardware should perform measurements to obtain the statistically significant expectation value via repeating the same quantum circuit execution for many times (e.g., the number of shots used for the execution on  the real quantum hardware in Section IV-B is 2000) resulting in  much more time consumption.

\section{Conclusion}
In this work, we investigate the accuracy and efficiency of the hybrid quantum-classical scheme for solving EFIE for analyzing electromagentic scattering from 3D PEC objects in CEM. Different from directly solving the original EFIE matrix equation system using the quantum algorithms, the hybrid scheme first designs the preconditioned linear system and then uses the double-layer iterative strategy for its solution, where the external
iteration layer builds subspace matrix equation systems with the smaller dimension and the internal iteration layer solves the smaller systems using the quantum algorithms.  The representative HHL and VQLS quantum algorithms executed on the quantum simulator and quantum computer platforms are considered in this work. The theoretical time complexity of the hybrid quantum-classical scheme is analyzed.  Numerical results are presented to demonstrate the accuracy and computational efficiency of the hybrid quantum-classical scheme for  the solution of EFIE. The results show that the computational complexity of the hybrid VQLS-classical scheme is lower than the conventional fast solvers in classical computing, which demonstrates the hybrid scheme is more promising for analyzing large-scale electromagnetic problems.

\end{document}